\documentclass[]{emulateapj}

\usepackage{color}

\newcommand{\SDSS}{SDSS-II}
\newcommand{\snana}{{\tt SNANA}}
\newcommand{\simlib}{{\tt SIMLIB}}
\newcommand{\unc}{uncertainty}
\newcommand{\uncs}{uncertainties}
\newcommand{\mlcs}{{\sc mlcs2k2}}
\newcommand{\SALTII}{{\sc salt--ii}}

\newcommand{\photoz}{photo-Z}
\newcommand{\zph}{Z_{\rm phot}}
\newcommand{\ztrue}{Z_{\rm true}}

\newcommand{\hostlib}{{\tt HOSTLIB}}
\newcommand{\spec}{spectroscopic}
\newcommand{\eff}{efficiency}
\newcommand{\ineff}{inefficiency}
\newcommand{\effspec}{\epsilon_{\rm spec}}

\newcommand{\NDATA}{N_{\rm DATA}}
\newcommand{\NSIM}{N_{\rm SIM}}
\newcommand{\modpar}{\vec{x}_{\rm model}}
\newcommand{\simpar}{\vec{x}_{\rm sim}}

\newcommand{\OM}{\Omega_{\rm M}}
\newcommand{\OL}{\Omega_{\Lambda}}

\newcommand{\dmfifteen}{\Delta m_{15}}

\newcommand{\wwwSNANA}{\tt http://www.sdss.org/supernova/SNANA.html}

\newcommand{\wwwMINUIT}{\tt http://wwwasdoc.web.cern.ch/wwwasdoc/minuit/minmain.html}
\newcommand{\wwwLSST}{\tt http://www.lsst.org/lsst}
\newcommand{\wwwDES}{\tt http://www.darkenergysurvey.org}

\newcommand{\wwwPDG}{\tt http://pdg.lbl.gov/2009/reviews/rpp2009-rev-probability.pdf}

\begin{document}

\title{ {\tt SNANA:} a Public Software Package for Supernova Analysis }

\submitted{Accepted for Publication in PASP}
\email{kessler@kicp.uchicago.edu}

\newcommand{\NUMUCASTRO}{1}  
\newcommand{\NUMKICP}{2}     
\newcommand{\NUMANL}{3}      
\newcommand{\NUMRUTGERS}{4}  
\newcommand{\NUMWAYNE}{5}    
\newcommand{\NUMFNAL}{6}     
\newcommand{\NUMSNT}{7}      
\newcommand{\NUMUPENN}{8}    
\newcommand{\NUMUW}{9}       

\author{
Richard~Kessler,\altaffilmark{\NUMUCASTRO,\NUMKICP}
Joseph~P.~Bernstein,\altaffilmark{\NUMANL}
David~Cinabro,\altaffilmark{\NUMWAYNE}
Benjamin~Dilday,\altaffilmark{\NUMRUTGERS}
Joshua~A.~Frieman,\altaffilmark{\NUMKICP,\NUMUCASTRO,\NUMFNAL}
Saurabh~Jha,\altaffilmark{\NUMRUTGERS}
Stephen~Kuhlmann,\altaffilmark{\NUMANL}
Gajus~Miknaitis,\altaffilmark{\NUMSNT,\NUMFNAL}
Masao~Sako,\altaffilmark{\NUMUPENN}
Matt~Taylor,\altaffilmark{\NUMWAYNE}
Jake~Vanderplas\altaffilmark{\NUMUW}
} 

\altaffiltext{\NUMUCASTRO}{
  Department of Astronomy and Astrophysics,
   The University of Chicago, 5640 South Ellis Avenue, Chicago, IL 60637
}

\altaffiltext{\NUMKICP}{
  Kavli Institute for Cosmological Physics, 
   The University of Chicago, 5640 South Ellis Avenue Chicago, IL 60637
}

\altaffiltext{\NUMANL}{
Argonne National Laboratory, 9700 S. Cass Avenue, Lemont, IL 60437
}

\altaffiltext{\NUMRUTGERS}{
Department of Physics and Astronomy, 
Rutgers University, 136 Frelinghuysen Road, Piscataway, NJ 08854
}

\altaffiltext{\NUMWAYNE}{
Department of Physics, 
Wayne State University, Detroit, MI 48202
}

\altaffiltext{\NUMFNAL}{
Center for Particle Astrophysics, 
  Fermi National Accelerator Laboratory, P.O. Box 500, Batavia, IL 60510
}

\altaffiltext{\NUMSNT}{
Center for Neighborhood Technology
2125 W. North Ave, Chicago IL 60647
}

\altaffiltext{\NUMUPENN}{
Department of Physics and Astronomy,
University of Pennsylvania, 203 South 33rd Street, Philadelphia, PA  19104
}

\altaffiltext{\NUMUW}{
  Department of Astronomy,
   University of Washington, Box 351580, Seattle, WA 98195
}

\begin{abstract}
We describe a general analysis package for supernova (SN) light curves, 
called \snana, that contains a simulation, light curve fitter,
and cosmology fitter.
The software is designed with the primary goal of using 
SNe~Ia as distance indicators for the determination of 
cosmological parameters, but it can also be used to study 
efficiencies for analyses of SN rates,
estimate contamination from non-Ia SNe,
and optimize future surveys.
Several SN models are available within the same 
software architecture, allowing technical features such
as K-corrections to be consistently used among multiple models,
and thus making it easier to make detailed comparisons
between models.
New and improved light-curve models can be easily added.
The software works with arbitrary surveys and telescopes
and has already been used by several collaborations, leading
to more robust and easy-to-use code.
This software is not intended as a final product release,
but rather it is designed to undergo continual improvements
from the community as more is learned about SNe.
We give an overview of the \snana\ capabilities,
as well as some of its limitations.
Interested users can find software downloads and
more detailed information from the manuals at {\wwwSNANA}.
\end{abstract}

\section{Introduction}
\label{sec:intro}

It has long been recognized that type Ia supernovae (SN~Ia) 
are a reasonably homogeneous class of events, 
with similar light curves, colors, and spectra.
After accounting for correlations between light-curve shape 
(usually the width or decline-rate), 
color, and luminosity, SNe~Ia provide relative distance modulus 
estimates with a dispersion of only $\sim 15$\%  \citep{Phillips_93}.
There are many approaches to characterizing the correlations 
that are used in fitting observed light curves,
and the most commonly used methods are shown in 
Table~\ref{tb:models}.
A typical model includes a combination of template light curves
and spectra, along with variable parameters (e.g., stretch and color)
that modify the templates to match the observed light curve.
The model  templates themselves are derived by a ``training'' 
procedure involving well-observed SNe.
Each light-curve fitting method has been implemented with different  
software, making comparisons between the models difficult. 
Comparison of light-curve  simulations based on the
different models is even more of a challenge, since there are no  
standard codes or methods.

The analysis of recent SN~Ia survey data has shown that the  
determination of cosmological parameters is now limited by 
systematic uncertainties \citep{Astier06,WV07,K09}
associated with the light-curve fitting methods. 
To make progress in overcoming these limitations, 
there is a need for reliable and reproducible analysis software 
that can be shared among different SN groups aiming to improve
or replace these methods. Even in cases where different groups use 
the same light-curve fitting method, they may use different 
versions of the software, again complicating comparisons.
Moreover, a comparison of different analysis methods can be complicated  
by subtle differences in analysis details such as the treatment 
of selection criteria and K-corrections. Unraveling
these differences calls for a single, well-documented analysis  
framework with clear software version control. 
In addition, such a framework can naturally incorporate  
detailed Monte Carlo simulations that are required to quantify 
the impact of survey selection efficiencies on  
light-curve fitting and cosmological inference.
Such simulations can also be used to investigate sources of 
intrinsic SN~Ia brightness fluctuations that result in 
scatter in the Hubble diagram beyond that caused by  
photometric errors.

To address these issues and thereby improve the availability, 
quality, and reproducibility of SN analyses,
we have developed a publicly available analysis software  
package called \snana. This package contains a light-curve fitter,
Monte Carlo simulation, and cosmology fitter. 
These tools incorporate existing SN light-curve models
(Table~\ref{tb:models}), and are easily adaptable to new
or improved models.
A description and comparison of the \mlcs\ and \SALTII\
methods is given in \S~5 of \citet{K09}.
The stretch-templates can represent Ia or non-Ia SNe, 
and can therefore be applied to any 
type of SN.

The \snana\ software package was originally developed 
and used for the analysis 
of the \SDSS\ SN sample \citep{Frieman07,K09}, 
but the need to plan for future SN surveys, such as the 
Dark Energy Survey (DES\footnote{\wwwDES}) and 
Large Synoptic Survey Telescope (LSST\footnote{\wwwLSST}), 
led to a publicly available package.
Additional motivation for public-use software is described in
\citet{AstroSoft09}.
This paper describes the capabilities of \snana,
which are rapidly evolving with the needs of larger
SN samples. A manual, installation instructions, and
the software are available from our website.\footnote{\wwwSNANA}

There are three basic principles in the design of \snana.
First, each light-curve model is defined in a separate
function that is used in both the light-curve fitter 
and in the simulation. This ensures that the fitter and
simulation are synchronized with the same underlying model,
and it also allows for anyone to provide a new SN model
by simply providing a new function. If the underlying model
describes SNe~Ia in the rest-frame, then a standard
K-correction utility is used based on the choice of
spectral template and filter bandpasses.
The underlying SN model can also describe SNe~Ia in the 
observer frame, and thus bypass the K-correction utility.
The second principle allows for the simulation of any
survey by using a standard ``observing conditions'' library
(\S~\ref{sec:sim}) that describes the cadence, seeing, 
atmospheric transparency, and readout noise.
The third principle is to allow a new survey to be added
without making code modifications.

\begin{table}[hb]
\caption{
  SN models available in \snana.        } 
\begin{center}
\begin{tabular}{ll}
\tableline\tableline
   model (reference)    & comment  \\
\tableline  
 \mlcs\ \citep{JRK07}  
             &  $UBVRI$ + K-corr \\
 \mlcs-IR  (this paper)
             &  $UBVRIYJHK$ + K-corr \\
 \SALTII\  \citep{Guy07} 
             & rest-frame flux vs. $\lambda$ \& epoch \\
 $\dmfifteen$ \citep{Prieto06} 
             &  $BVRI$ + K-correction \\
 stretch   \citep{Gold01}
             & rest-frame mags + K-corr \\
 two-stretch\tablenotemark{a} \citep{Hayden09}
             & rest-frame mags + K-corr \\
 non-Ia  (this paper) &  II(P,L,n), Ib, Ibc \\
\tableline  
\end{tabular}
\end{center}
\tablenotetext{a}{Two-stretch splits the template at peak brightness,
and stretches each part separately.}  
  \label{tb:models}
\end{table}

\section{Simulating SN Light Curves}
\label{sec:sim}

Ideally, survey simulations would be based on artificial 
SNe embedded into survey images at the pixel level.
While survey images are available within a collaboration,
they are typically not available to the community.
Even if survey images were available, full image-level 
simulations would require a large amount of computing to 
perform the many variations that are needed to
study systematic \uncs\ or to optimize a future survey.
We have instead developed a fast light-curve simulation
that is based upon actual survey conditions
and that therefore accounts for non-photometric conditions and
varying time intervals between observations due to bad weather.
At each survey epoch and sky location, the simulation uses 
the measured observing conditions to generate realistic
light curves that can be analyzed in exactly the same way as the data.

\begin{figure*}
\centering
\includegraphics[width=4.0in,angle=-90]{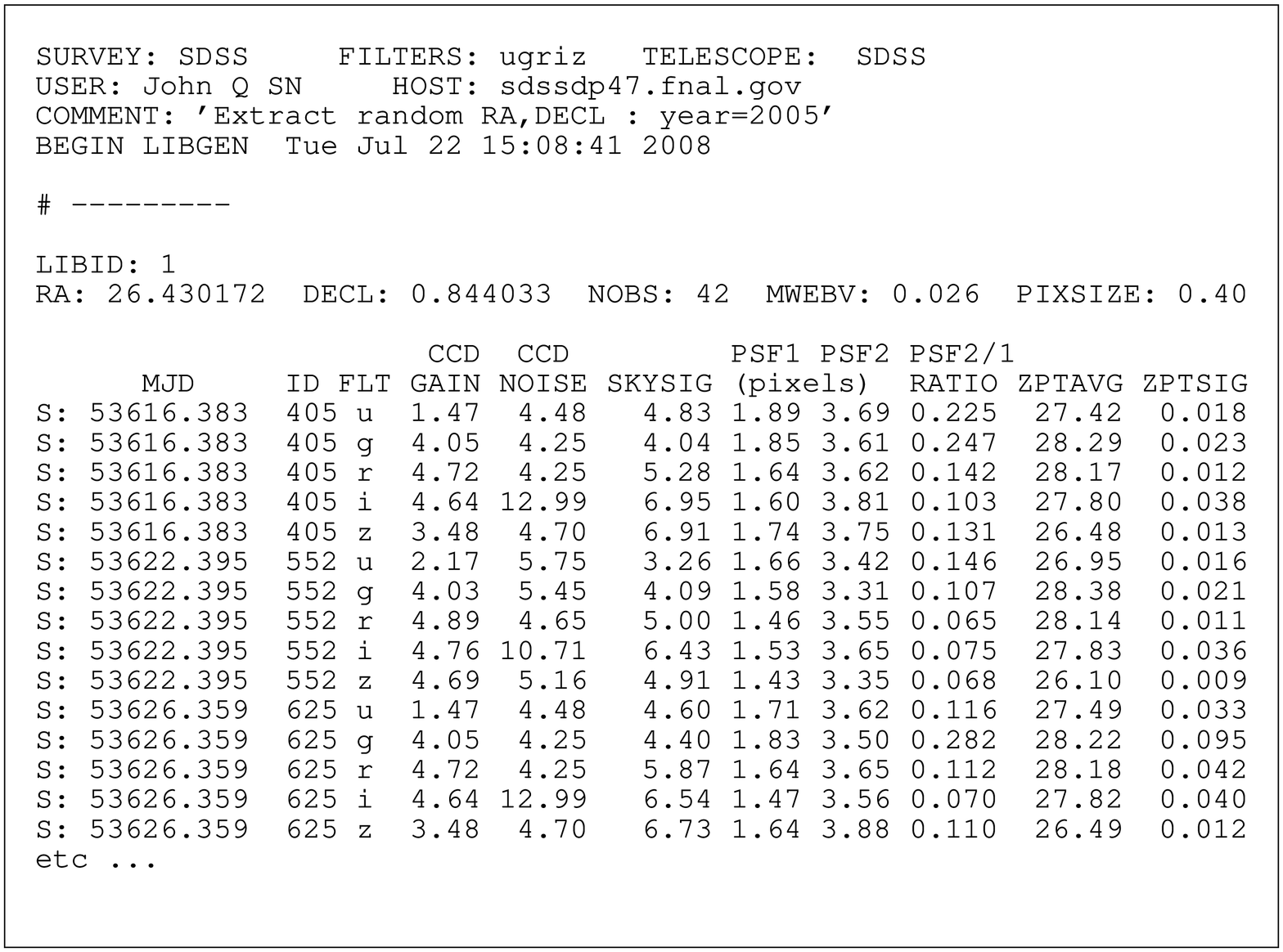}
  \caption{Example excerpt from a \simlib\ for the \SDSS\ SN Survey.}
\label{fig:simlib}
\end{figure*}

The observing conditions for a survey is stored in a text file
called a ``\simlib,'' an example of which is shown in
Fig.~\ref{fig:simlib}.  A simulated SN flux is generated
for each ``{\tt MJD}'' corresponding to an observation.
The quantity {\tt ZPTAVG} relates the magnitude ($m$) 
to observed flux ($F$) in CCD counts by the relation
$F = 10^{-0.4(m-{\tt ZPTAVG})}$; {\tt ZPTAVG} therefore 
accounts for the atmospheric transparency, exposure time, 
and the \eff\ of the telescope, CCDs, and filters.
To optimize a future survey, the exposure time can be 
independently varied to avoid re-making the \simlib\
with different {\tt ZPTAVG} values.


The CCD readout noise and sky-noise are determined by the 
{\tt CDD-NOISE} and {\tt SKYSIG} parameters (per pixel) 
summed in quadrature over an effective aperture 
($A$) based on PSF-fitting,
\begin{equation} 
 A = 2\pi\frac{ \left[ \int PSF(r)rdr \right]^2}{\int [PSF(r)]^2 r dr}~.
\end{equation}
For a PSF described by a Gaussian 
($\sigma$ = {\tt PSF1} in Fig.~\ref{fig:simlib}), the effective aperture
is $4\pi\sigma^2$. The simulation accounts for non-Guassian
tails in the PSF by including a second Gaussian function
parametrized by {\tt PSF2} and {\tt PSF2/1-RATIO}.
For the \SDSS\ SN simulation, the inclusion of non-Gaussian
tails increases the total noise by about 10\% compared
to the single-Gaussian approximation for the PSF.
The quantity {\tt ZPTSIG} reflects the spread in zeropoints
from an ensemble of standard stars, and is used to introduce
additional photometric dispersion.

For a small survey field, the \simlib\ needs
just one sequence of observations since the conditions within a
small field are likely to be uniform. For a large field, 
such as the $300$ square degree field of 
stripe 82 in the \SDSS\ SN Survey \citep{Frieman07}, 
a \simlib\ can contain thousands of observation sequences to account
for the variations across the field.
Whether SN collaborations use \snana\ or not, it would be useful
if all SN survey teams create appropriate \simlib\ files and
make them available along with data releases.
The current \snana\ package includes {\simlib s} for the 
\SDSS\ \citep{K09}, ESSENCE \citep{WV07}, SNLS \citep{Astier06}, 
and HST \citep{Riess_06} surveys,
as well as estimates for future surveys including the 
DES \citep{DES-moriond2009} and LSST surveys.
For completed SN surveys, the generation of a \simlib\ 
is straightforward to extract from the observations.
To study future surveys, the \simlib\ generation is 
typically based on extrapolating previous observations 
at a similar site, and including observing gaps due to 
the moon and telescope operations.
The reliability in predicting a \simlib\ for a future survey
is crucial to the reliability of the simulation results.

The simulation begins with the generation of SN properties
based on the selected model from Table~\ref{tb:models},
typically choosing a shape-luminosity and color parameter.
The color parameter can represent host-galaxy extinction ($A_V$)
for models such as \mlcs,
or it can reflect an agnostic relation between the supernova color
and its peak brightness (i.e, for models like {\SALTII}).
The parent distributions of these parameters are currently
based on a nearly complete sample of \SDSS\ SNe as described in 
\S 7 of \citet{K09}. 
The shape and color parameters, along with other internal
model (training) parameters such as light curve templates
and spectral surfaces, are used by the light curve model to 
determine the rest-frame magnitude at each epoch. 
To study potential biases from evolving SN properties,
an arbitrary redshift dependence can be assigned to 
the parent distributions of the shape and color parameters,
as well as to the model parameters that relate shape and color to 
intrinsic brightness 
(e.g., \SALTII\ $\alpha,\beta$, and host-galaxy dust parameter $R_V$).
The simulation also includes ``intrinsic smearing'' options to 
introduce additional scatter in the Hubble diagram residuals.

\begin{figure*}  
\centering
\includegraphics[width=1.6in]{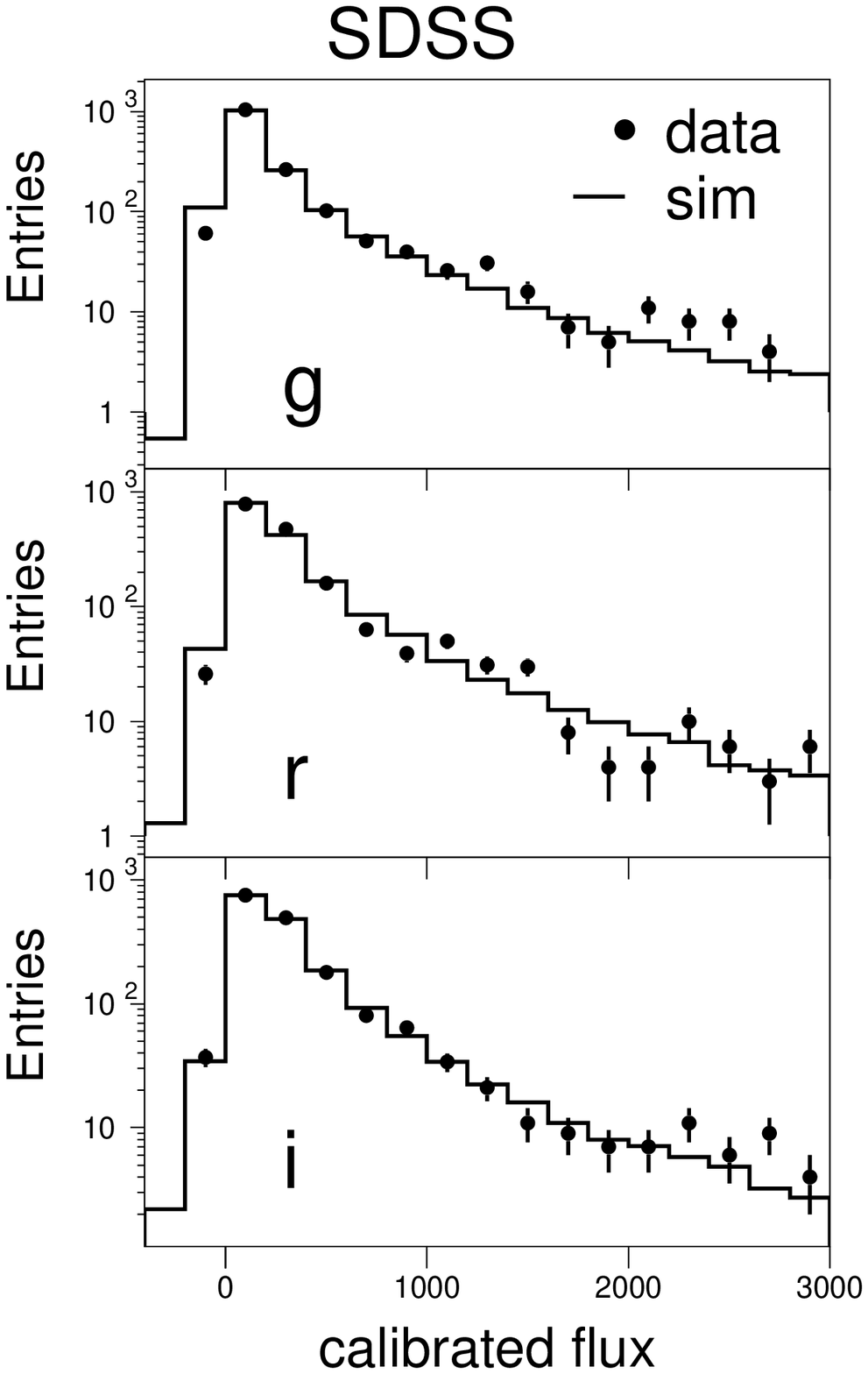}
\includegraphics[width=1.6in]{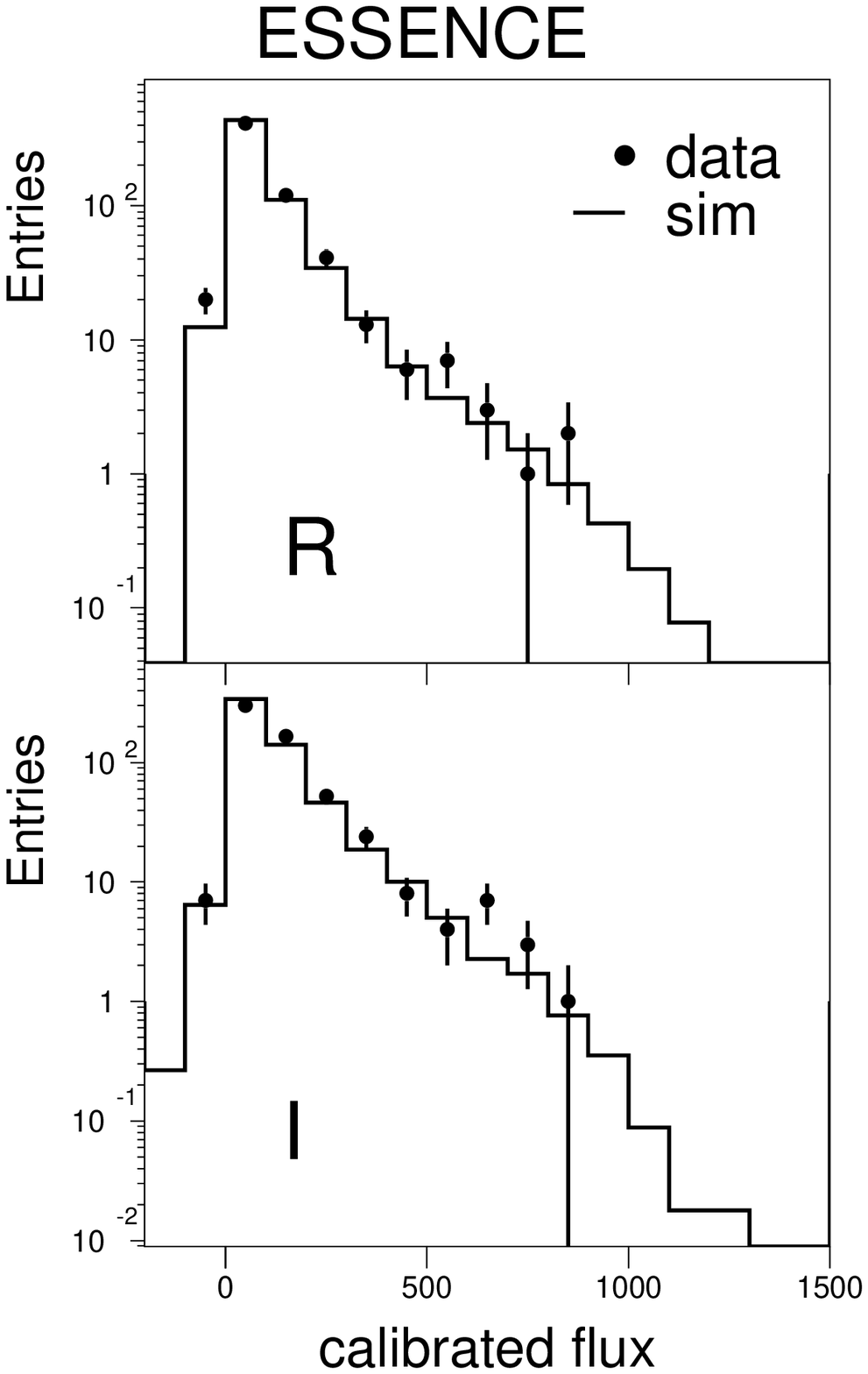}
\includegraphics[width=1.6in]{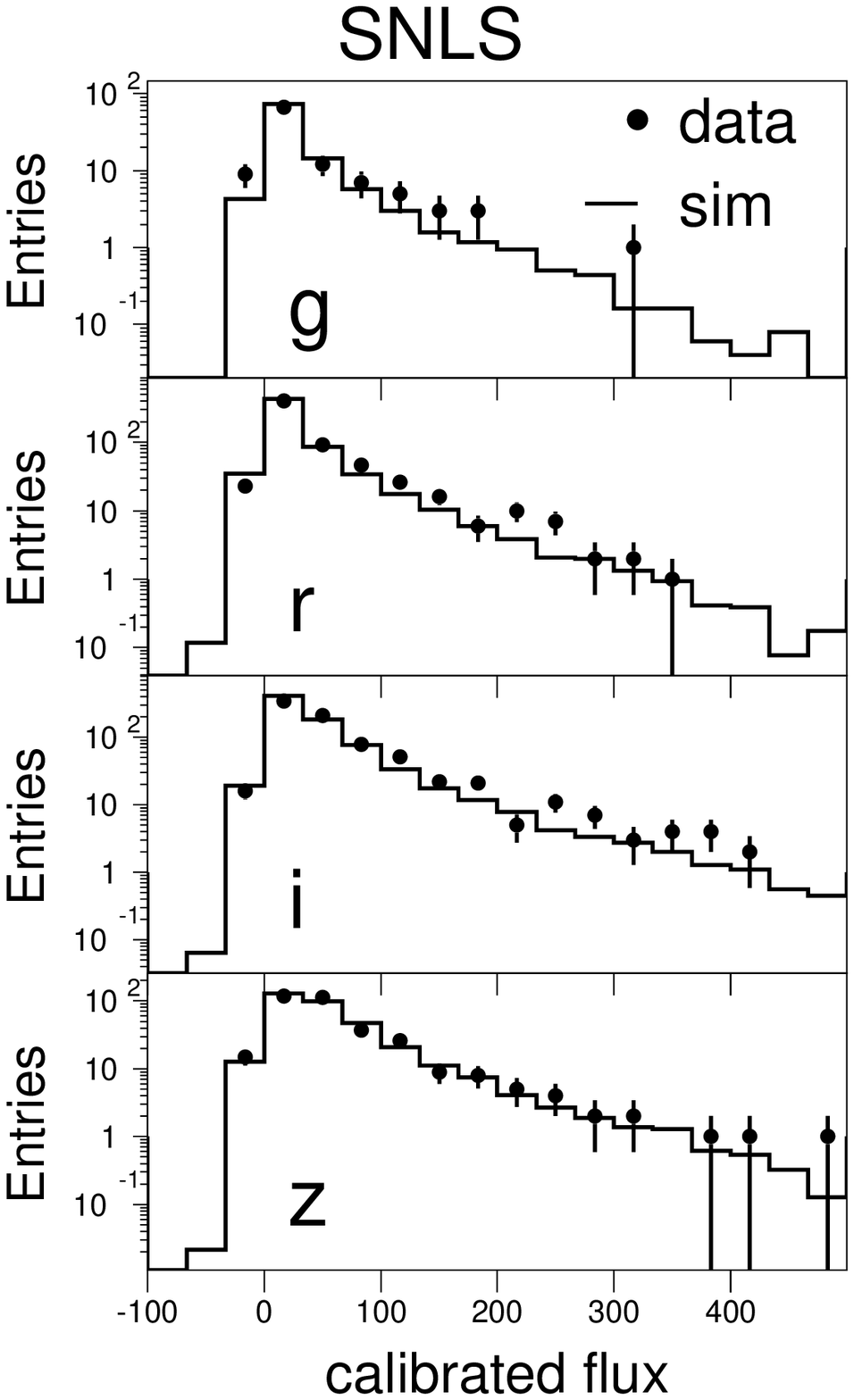}
\includegraphics[width=1.6in]{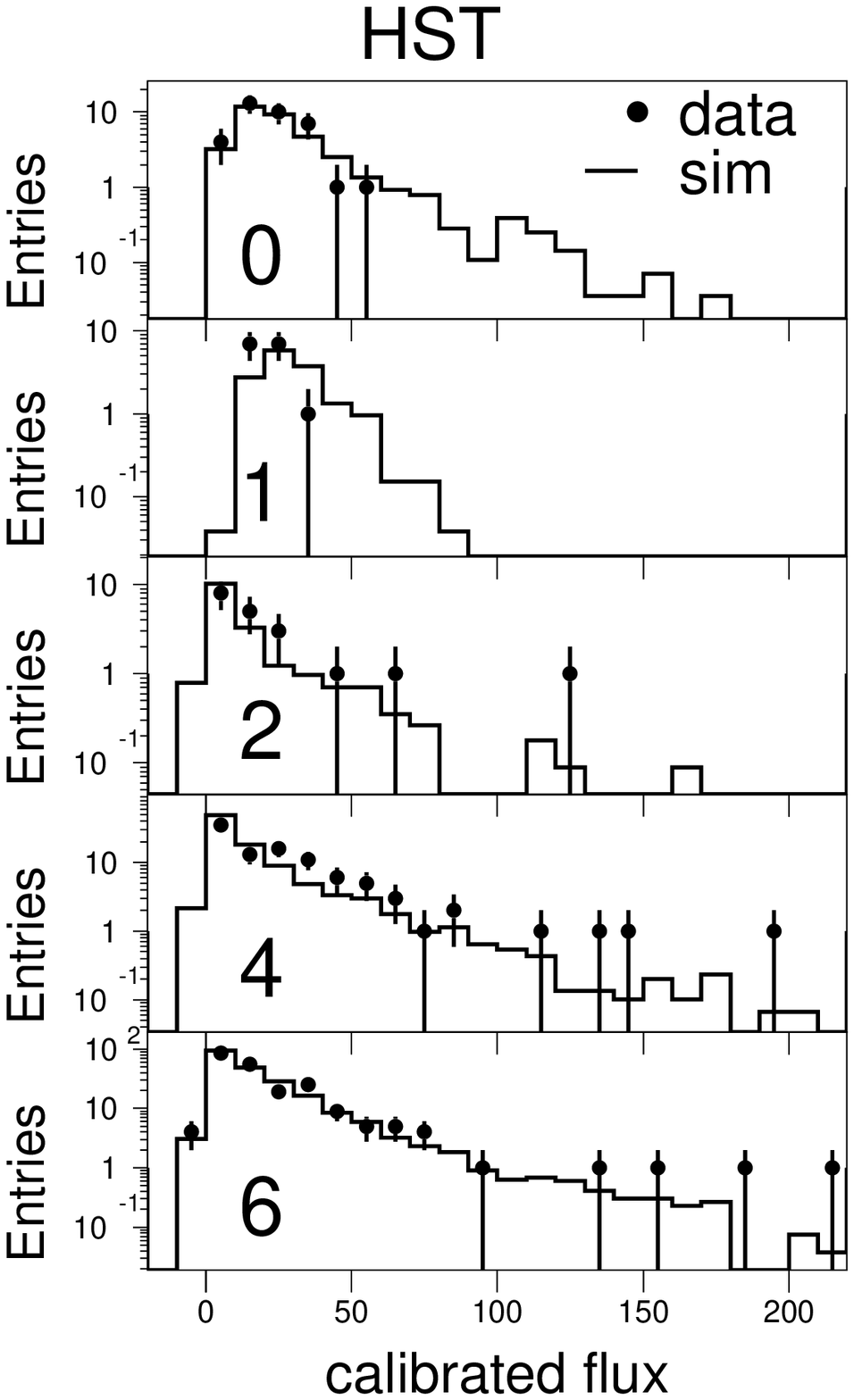}
  \caption{
	Comparison of flux distributions for data (dots) and 
   	simulations (histograms) for 
   	\SDSS, ESSENCE, SNLS, and HST.
   	The observer-frame passband is indicated on each plot
  	(HST filter symbols are
   	$0=$NIC-F110W, $1=$NIC-F160W, $2=$ACS-F606W, 
   	$4=$ACS-F775W, $6=$ACS-F850LP).
   	Each simulated distribution is scaled to have
   	the same total number of entries as the data. 
   	The calibrated flux is $10^{-0.4\cdot {\rm mag} + 11}$.
      }
  \label{fig:ovdatasim_flux}
\end{figure*}

\begin{figure*}  
\centering
\includegraphics[width=1.6in]{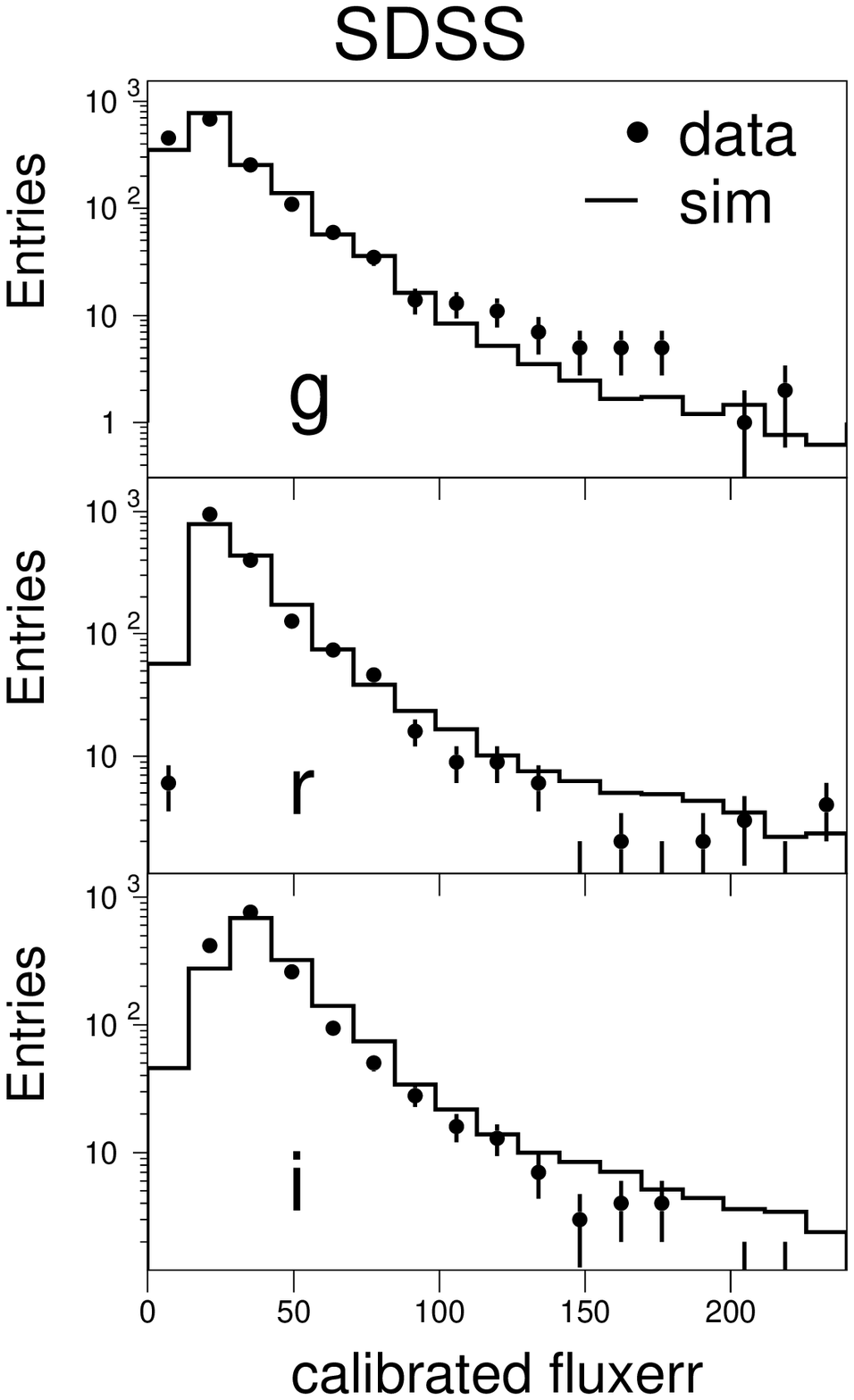}
\includegraphics[width=1.6in]{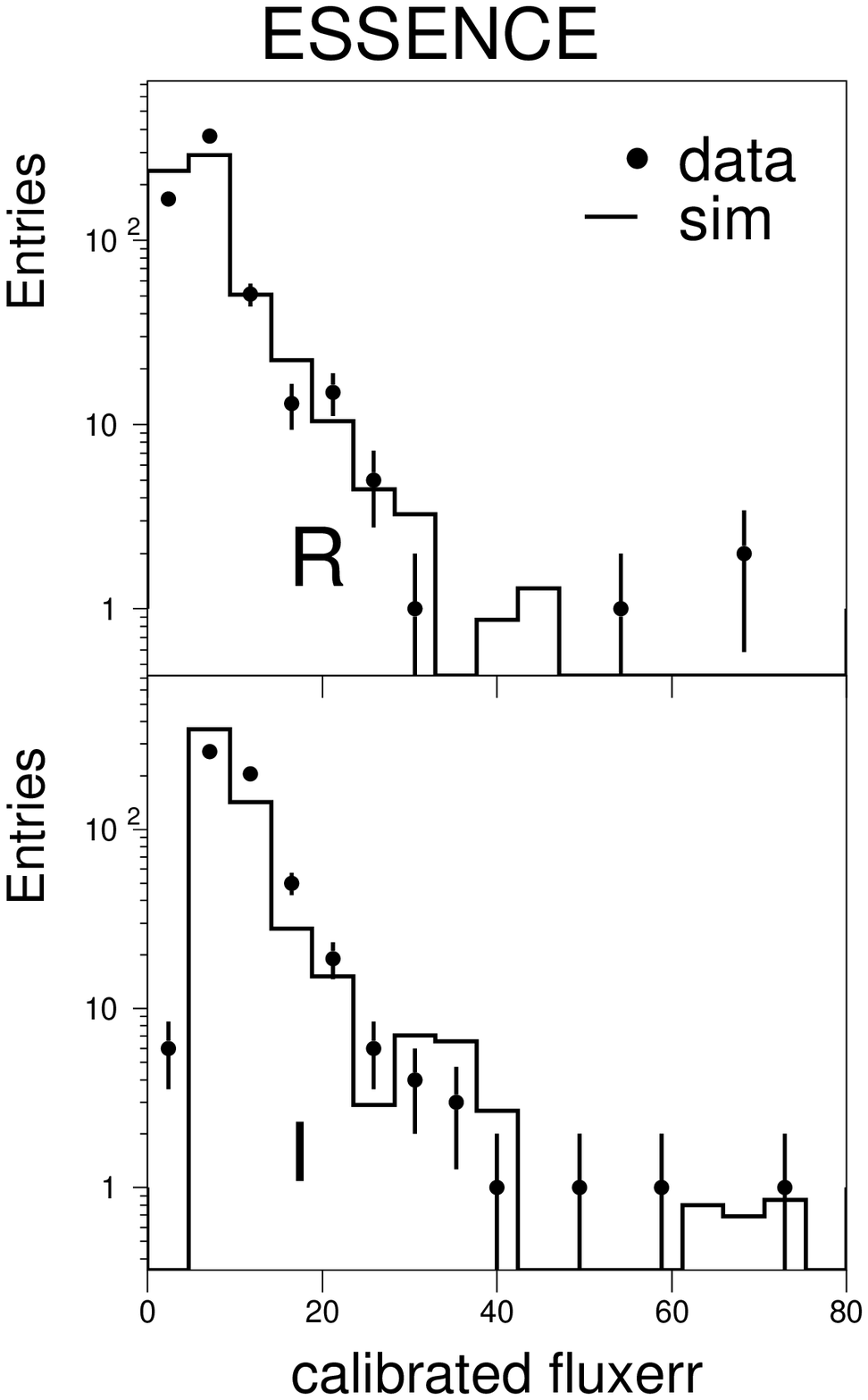}
\includegraphics[width=1.6in]{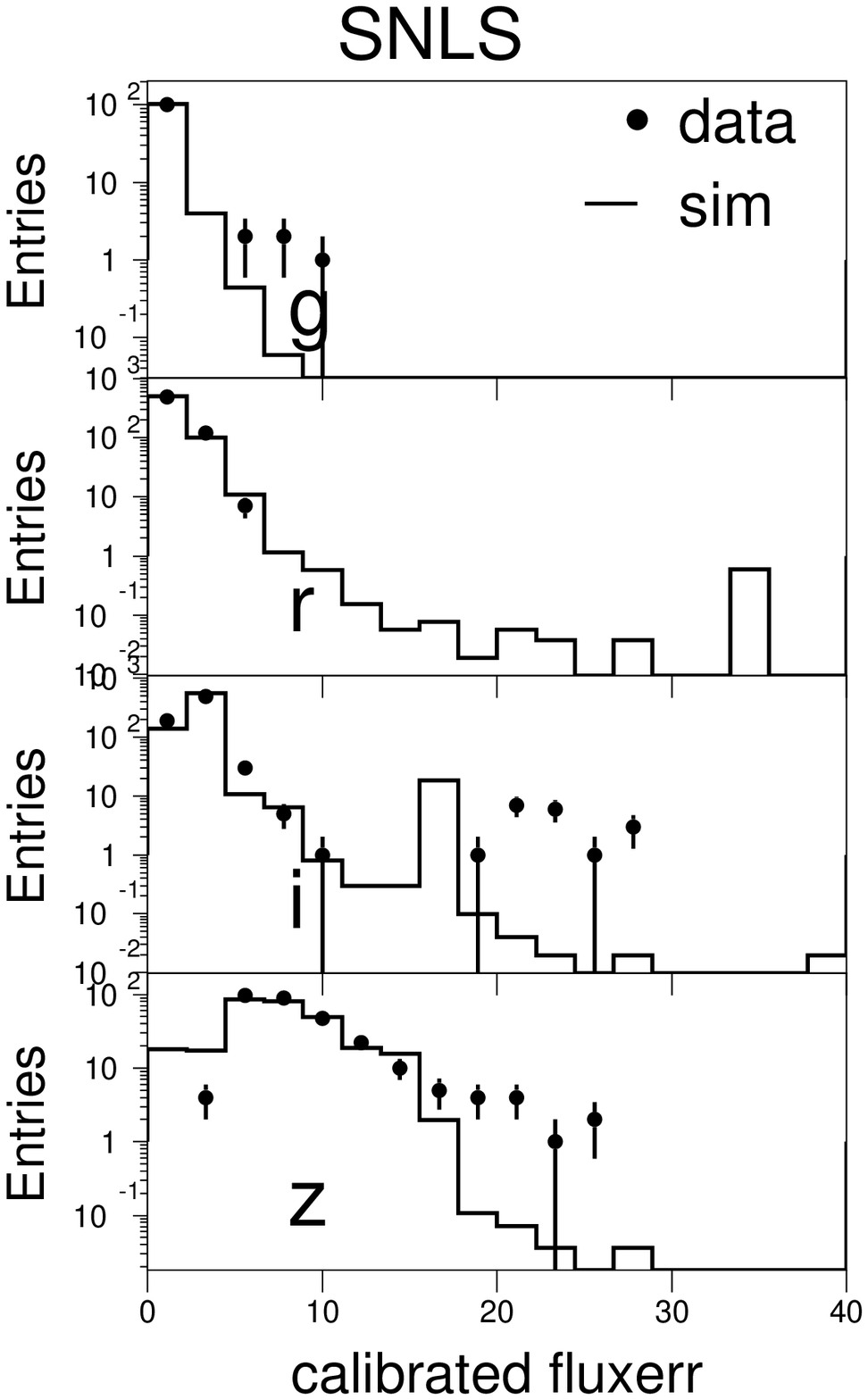}
\includegraphics[width=1.6in]{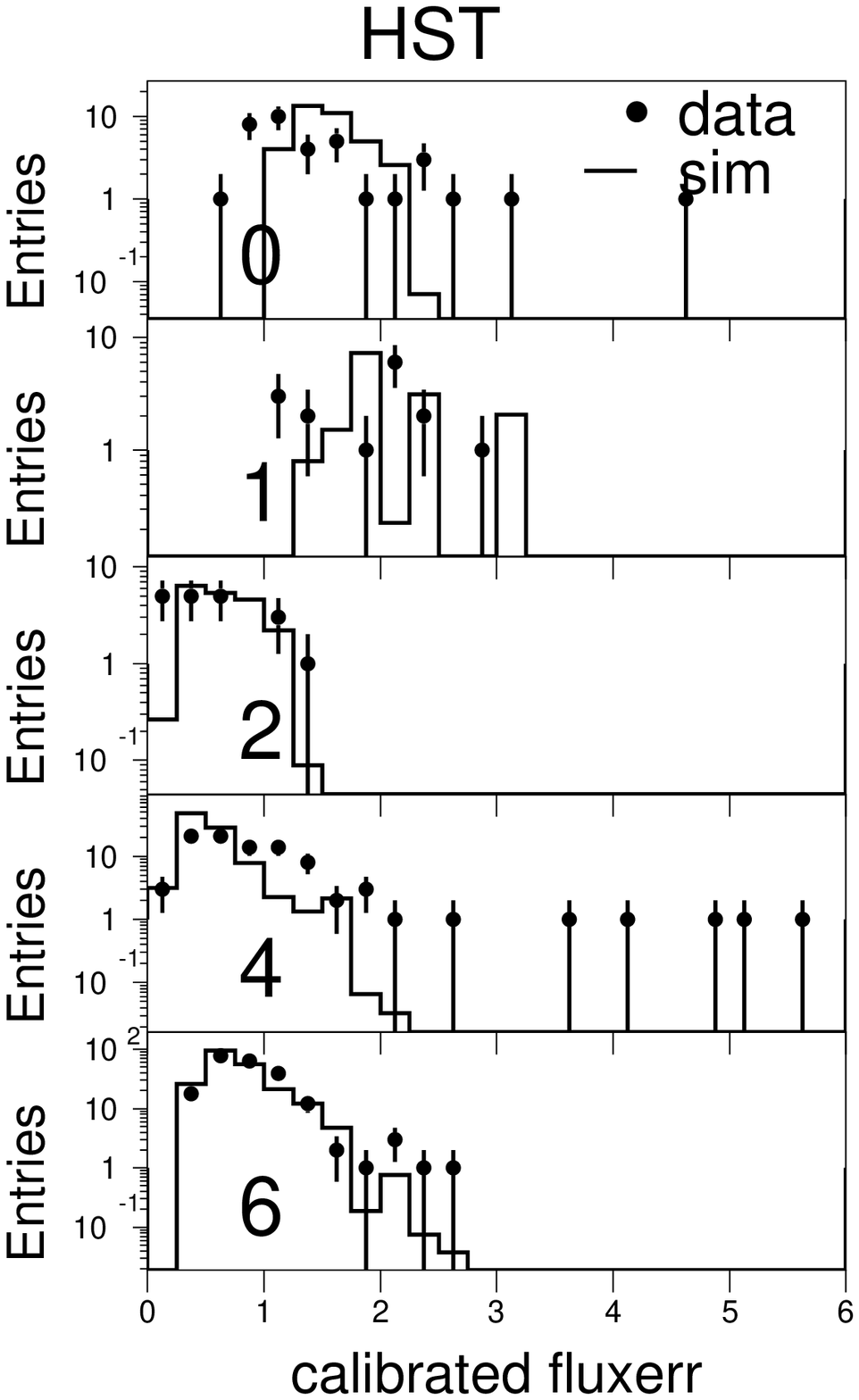}
  \caption{
	Data-simulation comparison of the flux \unc.
	See Fig.~\ref{fig:ovdatasim_flux} caption for details.
      }
  \label{fig:ovdatasim_fluxerr}
\end{figure*}

K-corrections similar to the method in \citet{Nugent2002}
are used to transform rest-frame model magnitudes to 
observer-frame magnitudes in the telescope system.
For most applications the spectral templates from
\citet{Hsiao07} are used to determine the K-corrections,
but the \snana\ software allows for using any spectral
template.
For the final step in the simulation,
the \simlib\ is used to translate the ideal above-atmosphere 
magnitudes into observed fluxes and \uncs.
As explained in \S6 of \citet{K09}, the simulation for each survey 
accounts for search-related efficiencies; in particular,
the \spec\ \eff\ is based on the discrepancy between the observed
and simulated redshift distribution.
The quality and flexibility of the simulation is illustrated
in Figs.~\ref{fig:ovdatasim_flux}-\ref{fig:ovdatasim_fluxerr},
which show data-simulation comparisons for the measured flux
and its \unc\ for four different surveys.
For the flux-\unc\ comparison (Fig.~\ref{fig:ovdatasim_fluxerr}),
there are some discrepancies that may be due to a 
mis-interpretation of the noise determination from the
image-subtraction pipeline.

Although supernova-based cosmology results have all used
rest-frame SN~Ia models of optical ($UBVRI$) light curves,
there is a growing interest in obtaining infrared  light curves
in order to reduce the effect of host-galaxy extinction.
To study such possibilities in future surveys,
rest-frame infrared light curves can be simulated 
using an extended version of \mlcs\ in which 
light curve templates for $YJHK$ have been constructed based 
on recent observations from PAIRITEL \citep{WV08IR}.

To study the determination of SN~Ia photometric redshifts ({\photoz}), 
the simulation includes a host-galaxy \photoz\ library
({\hostlib}) so that the host-galaxy \photoz\ can be used as a 
prior in a \photoz\ fit.  
The \hostlib\ is not created by \snana, and therefore the \hostlib\
must be created externally from either another simulation,
or from a \photoz\ analysis of host-galaxies with \spec\ redshifts.
The \snana\ simulation simply finds a host galaxy with a redshift 
matching the SN redshift, and stores the host-galaxy \photoz\ 
to be used by the light curve fitter.

To estimate SN~Ia sample contamination,
the simulation can generate non-Ia SN light curves.
While the Type Ia light curve models are based on
a parametric equation or photometric templates, 
there are no such models for non-Ia types.
To allow for a large diversity of non-Ia SNe, the simulation
uses a library of spectral templates that specify the
SN flux as a function of epoch and wavelength.
A spectral template can be a smoothed average based on an 
ensemble of SNe, or a template can correspond to 
a particular (well-observed) supernova where a composite 
spectrum is warped to match the observed photometric colors.
Since the spectral templates are interpolated to cover all wavelengths
and epochs, only well-sampled photometric light curves are used.
The current library contains composite and individual
templates for types IIP, IIn, Ib, and Ibc.
The simulation of the non-Ia diversity will gradually improve
as more well-observed non-Ia light curves become available.
A limitation of \snana\ is that there is currently no 
utility to convert non-Ia light curves into spectral surfaces.

A trigger simulation is included to mimic algorithms 
that are used to discover SNe~Ia in real time.
Although the offline selection criteria should in 
principle be more stringent than the search trigger,
it may be useful to use the simulation to make more
rigorous crosschecks.  The trigger simulation is also
useful to anticipate SN discovery rates in the design of 
future surveys. 
The trigger simulation is based on user-supplied \eff\ 
tables that depend either on observed magnitudes or on the 
signal-to-noise ratio (SNR).
These \eff\ tables could be made, for example, by injecting
fake SNe into real images, and then passing these images
through the object-finding pipeline.
Since a discovery typically requires observations in 
multiple filters and/or multiple epochs, 
one can specify arbitrary detection logic.
For example, the \SDSS\ SN Survey required
a detection in at least two of the $gri$ filters,
and required a detection at two or more epochs;
this logic is simulated by specifying the trigger
to be ``2  gr+ri+gi.''

A more difficult aspect of the trigger is the
target selection for \spec\ observations.
Since \spec\ selection is based on human decisions and 
telescope availability, there is no clear software algorithm 
to describe this part of the \eff. 
As described in \S~6.2.1 of \citet{K09},
the observed and simulated redshift distributions
can be used to empirically determine the
\spec\ selection \eff. 
After all quantifiable sources of {\ineff} are included
in the simulation, the simulated redshift distribution will
typically not match that of the data. In the specific case
of \citet{K09}, the observed redshift distribution always
has a deficit at higher redshifts compared to the simulation, 
and this deficit is assumed to be a result of the 
\spec\ \eff\ ($\effspec$).
In general, $\effspec$ depends on two types of parameters. 
The first set of parameters are those determined by the simulation
(denoted $\simpar$), such as intrinsic or observed brightness,
host-galaxy extinction, and host-SN separation.
The second set of parameters (denoted $\modpar$) 
are those introduced by the analysis team based on their 
knowledge of survey operations, or in some cases based on a guess.
A simple example is $\effspec = \exp[-(r-r_0)/\tau_r]$,
where $\simpar=r$ is the observed $r$-band magnitude,
and $\modpar = \{ r_0,\tau_r \}$ are the model parameters. 
The $\modpar$ parameters are determined by solving
\begin{eqnarray}
&&    \int d\simpar 
   \left[ {\NSIM}(z_i,\simpar) \effspec(z_i,\simpar,\modpar) \right] 
\nonumber \\
&&      =  {\NDATA}(z_i)~,
   \label{eq:simpar}
\end{eqnarray}
where $\NDATA(z_i)$ is the number of observed SNe, 
after all selection criteria, 
in the redshift bin centered at $z_i$, 
and $\NSIM(z_i,\simpar)$ is the number of simulated SNe
in the same redshift bin and within the integration cell denoted
by $d\simpar$.
\snana\ can be used to determine 
$N_{\rm SIM}(z,\simpar)$ and  $N_{\rm DATA}(z)$,
but the extraction of $\modpar$ requires specific software 
that is currently not part of \snana.
It may be convenient to determine the model parameters
in redshfit bins, $\modpar(z_i)$, and then fit each parameter
to a smooth function of redshift.
With each new definition of $\modpar$ parameters, the \snana\ simulation
is easily modified to compute $\effspec(z,\simpar,\modpar)$.
Finally, we note that the precision in determining $\effspec$
is limited by the precision in the SN rate as a function of
redshift.

\section{Light Curve Fitting}
\label{sec:fit}

The light curve fitting program is appended to an underlying 
program that reads SN light curves and applies user-defined 
criteria to select a subset of SNe and epochs.  
Users can append other fitters or other analysis programs
to the underlying program. This structure ensures a uniform definition 
of selection requirements that can be shared among users.

\begin{figure}[hb]
\centering
 \epsscale{1.15}
 \plotone{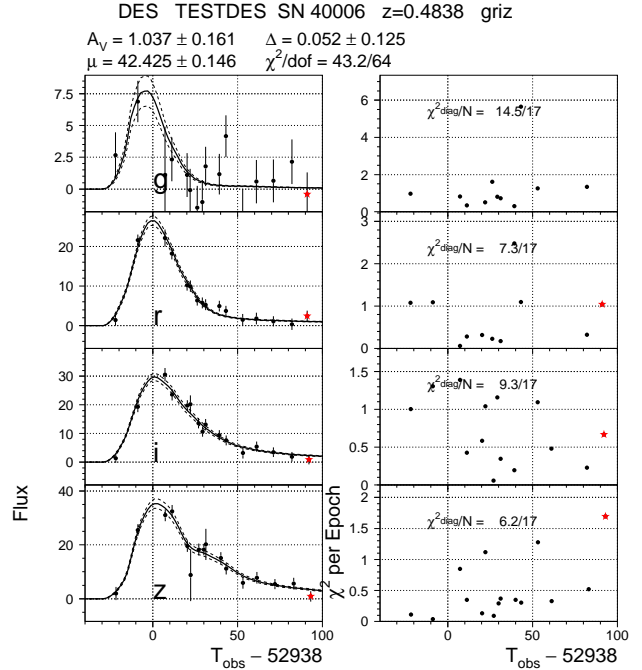}
  \caption{
    	As part of a study to optimize the DES-SN Survey,
	\snana-\mlcs\ light curve fits (left) are shown 
	for a simulated SN~Ia at redshift $z=0.48$ with host-galaxy
	extinction $A_V \sim 1$.
	The observer filter ($g,r,i,z$) is indicated on each panel.
	Solid dots with error bars are the measurements and \uncs,
	stars indicates measurements outside the specified epoch range,
	solid curve is the best-fit model, and
	dashed curve indicates the model \unc. 
	The right panels show the residual $\chi^2$ maps 
	that are useful for identifying outliers.
      }
  \label{fig:snfit_DES}
\end{figure}

\begin{figure}[hb]
\centering
\epsscale{1.05}
\plottwo{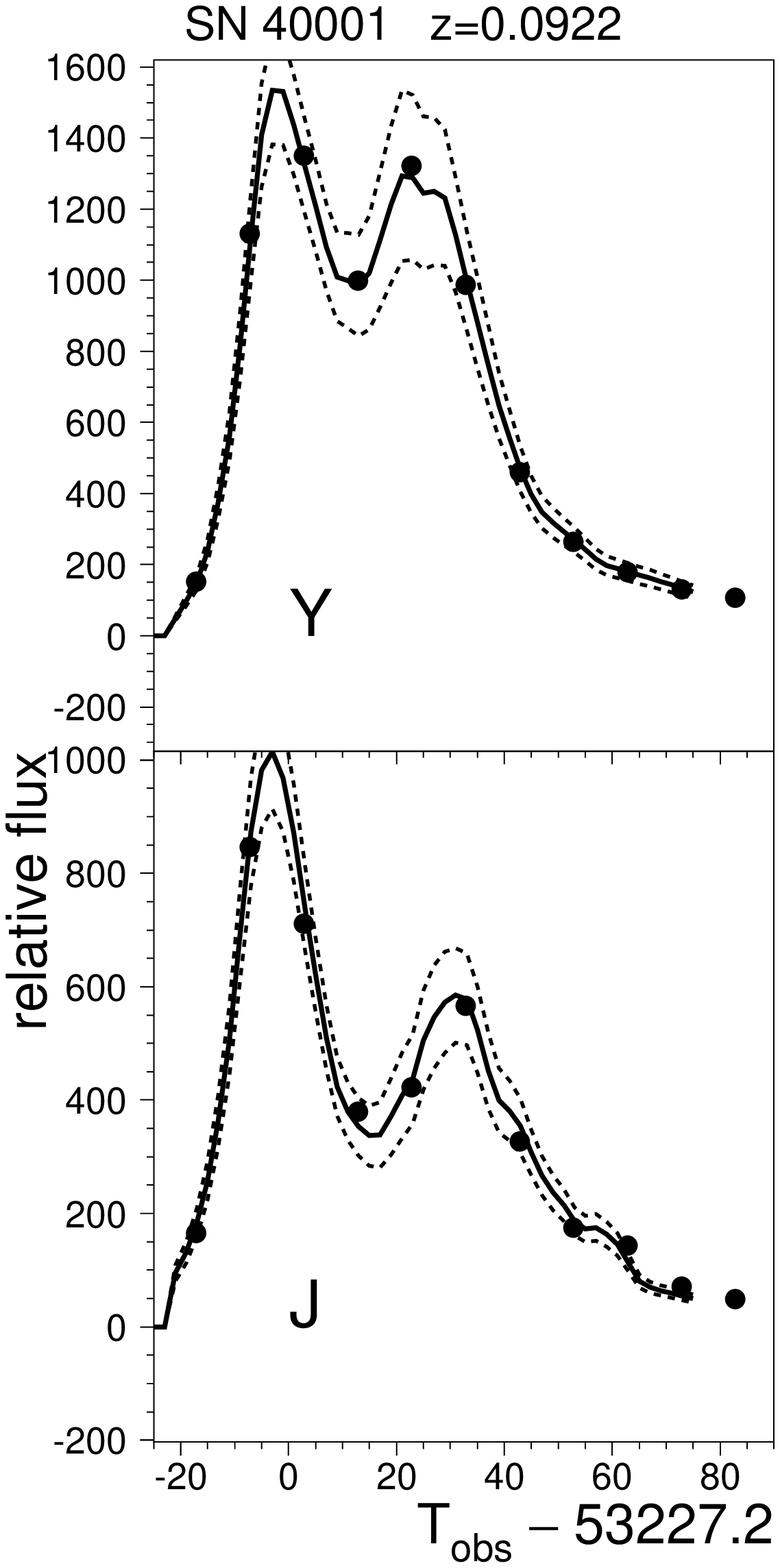}{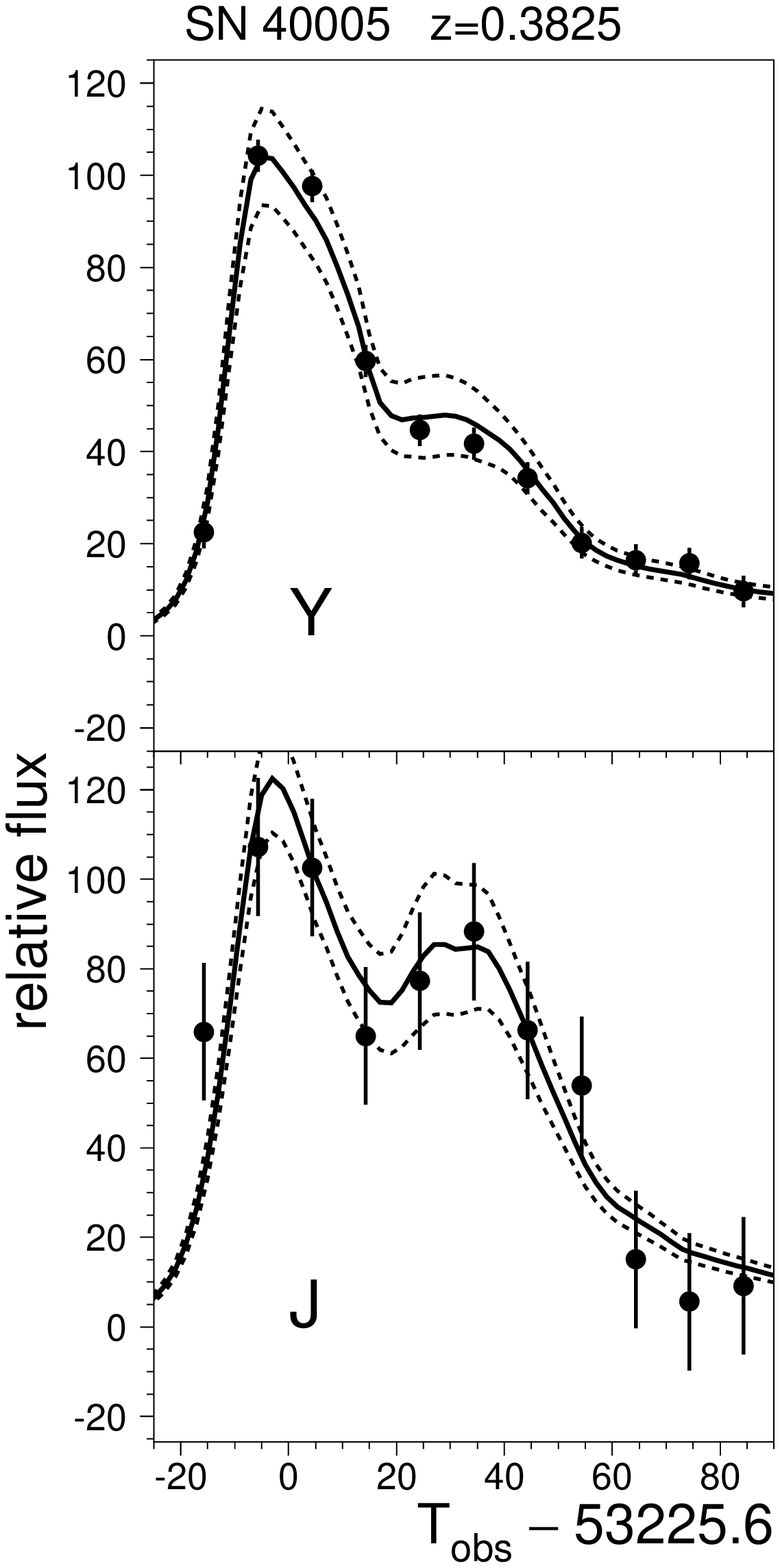}
  \caption{
    	As part of a study to observe DES-supernova fields 
	in the infrared with VISTA 
	(Visible and Infrared Survey Telescope for Astronomy:
	{\tt www.vista.ac.uk}),
	\snana-\mlcs\ light curve fits are shown 
	for simulated SNe~Ia at redshifts shown above each panel.
	The observer filter ($Y,J$) is indicated on each panel.
	Symbols and curves are defined in the
	Fig.~\ref{fig:snfit_DES} caption.
      }
  \label{fig:snfit_VIDEO}
\end{figure}

The \snana\ light curve fitter treats data and
simulated samples in exactly the same way.
Available models are shown in Table~\ref{tb:models}.
Models that describe SNe in the rest-frame all use the 
same algorithms for host-galaxy extinction and K-corrections.
The standard fit assumes a well-known \spec\ redshift,
and then fits for shape and color parameters.
For the \mlcs\ model, each fit returns four parameters:
epoch of peak brightness ($t_0$),
shape-parameter ($\Delta$),
host-galaxy extinction ($A_V$),
and a distance modulus ($\mu$).
For other similar models, the shape-parameter $\Delta$ is 
replaced by the corresponding parameter, such as stretch
or $\dmfifteen$.
Observer-frame infrared light curves ($YJHK$, HST, etc ...) 
can be fit with any model at redshifts that correspond to 
the rest-frame optical region of the model. 
For \mlcs, the rest-frame model includes $YJHK$.
Example light curve fits in the optical and infrared 
(using the \mlcs\ model) are shown in 
Figs.~\ref{fig:snfit_DES}-\ref{fig:snfit_VIDEO}.

\begin{figure}[hb]
 \centering
 \epsscale{1.1}
 \plotone{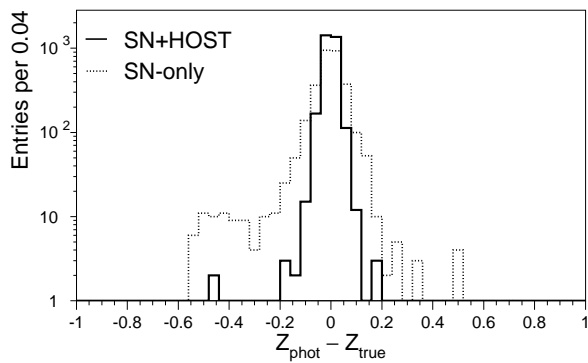}
  \caption{
	Distribution of $\zph - Z_{\rm true}$
    	from a study of the DES-SN Survey in a scenario
	in which \spec\ redshifts are not available.
	$\zph$ is the SN photometric redshift obtained
	with the \mlcs\ method.
	Dashed histogram is based solely on the SN photometric
	light curve, and the solid histogram includes the host-galaxy
	\photoz\ as a prior.
      }
  \label{fig:DES_photoz_resid}
\end{figure}

For surveys that cannot obtain a spectrum for all SNe,
there are two \photoz\ fit options.
The \mlcs\ model is used here for illustration.
The first method, called a ``constrained \photoz\ fit,'' 
is designed to identify SN~Ia that do not have
a \spec\ redshift: uses include SN rate measurements and 
targeting host-galaxy redshifts for unconfirmed SNe~Ia.
In this 4-parameter fit, the redshift is a fitted parameter 
and the distance modulus is constrained (calculated)
so that the observed luminosity of the SN corresponds
to a particular cosmology: $\mu = \mu(z,\OM,\OL,w)$
where $\OM,\OL,w$ are fixed by the user.
The second method, called a ``cosmology-photoZ'' fit,
involves fitting for both the redshift and distance modulus:
the five fitted parameters are
$\mu$, $z$, $t_0$, $\Delta$, $A_V$.
A potential difficulty with the 5-parameter 
cosmology-photoZ fit is CPU time:
the marginalization (see \S~\ref{sec:soft}) 
takes a few minutes per fit,
so studying a sample of $10^4$ SNe~Ia requires about 
a CPU-month of resources.\footnote{Processing times
are based on an Intel i686 2.8 GHz processor.}
In addition to these \photoz\ methods, the host-galaxy \photoz\  
can be used as a prior to help constrain the redshift.
For simulated SNe~Ia in the Dark Energy Survey (DES),
Fig.~\ref{fig:DES_photoz_resid} shows the difference between
the fitted \photoz\ ($\zph$) and true redshift ($\ztrue$).
As expected, including the host-galaxy \photoz\ as a prior 
results in a smaller scatter in $\zph - \ztrue$.

All fitting methods use {\sc minuit}\footnote{\wwwMINUIT} 
to obtain parameter values that minimize the data-model $\chi^2$.
There is also an option to obtain marginalized parameter
estimates\footnote{\wwwPDG}
at the expense of taking more CPU time, 
particularly for the 5-parameter cosmology-photoZ fit.
The interaction between fitting and selection requirements
introduces a subtle caveat in selecting measurements within
a specific epoch range relative to the time ($t_0$)
of maximum brightness. The epoch range is not well defined
before the fit, making the epoch selection ambiguous.
The \snana\ light curve fitter handles this situation by
performing two fit iterations. 
In the first iteration, the epoch range is very wide.
The first estimate of $t_0$ is then used to define the 
epoch-window for the second iteration.
The star in Fig.~\ref{fig:snfit_DES} indicates
a measurement that was rejected in the second iteration
because it is outside the user-specified epoch window.

The light-curve fitter includes the capability to 
determine interpolated fluxes and uncertainties at any epoch.
The uncertainty accounts for covariances among the model parameters.
This interpolation is useful to compare observed magnitudes
with synthetic magnitudes based on spectra, 
and to compare magnitudes from multiple telescopes
that observe the same SN at different times.

\section{Cosmology Fitter}
\label{sec:cosmofit}

\snana\ includes a basic cosmology fitter
that reads a table of redshifts and distance moduli
generated by the light curve fitter, 
and then determines cosmological parameters
$w$ and $\OM$ within the framework of the
Friedmann-Robertson-Walker model.
Priors on the cosmology parameters can be included from 
measurements of Baryon Acoustic Oscillations 
and the Cosmic Microwave Background.
Covariances from peculiar velocities \citep{Hui_06}
can be incorporated into the fitter,
which effectively downweights the most nearby SNe.

To study more exotic cosmology models, such as time-dependent
dark energy and non-zero  curvature, 
there is another fitter based on Monte Carlo Markov Chains.

\section{Software Overview}
\label{sec:soft}

The \snana\ software is written primarily in 
C and fortran, and makes extensive use of 
the CERN library ({\tt CERNLIB}) for fitting, 
creating plots, and for a graphical interface. 
For analysis features that appear to be specific to a 
particular data sample or light curve model, 
there is always an effort to integrate such features in a 
general manner that can be used more broadly.
For example, the use of priors and marginalization that
are associated with the \mlcs\ method have been 
integrated into \snana\ so that they can be used
on any light curve model.
In addition to the \snana\ simulation and fitters, 
there is an increasing library of software tools for 
comparing results, translating formats, plotting,
sorting, generating a \simlib, and converting $\chi^2$ 
grids into contours.
Although the code is stored in a CVS repository
for the developers,
the public code is released in incremental versions
(e.g., {\tt v7\_07}) with release notes documenting the changes.

In anticipation of very large SN surveys in the near future,
significant attention has been paid to processing speed.
The simulation generates more than 10 light curves per second.
The light curve fitter takes about 1 second per light curve
for the minimization, but the marginalization can take
10-100 seconds depending on the number of fit parameters
and the number of integration bins.

\section{Conclusion}
\label{sec:conclude}

There are currently about a dozen SN-related analyses that 
make use of \snana, and users include students, postdocs,
and faculty.
This broad usage has greatly helped to improve the 
reliability and usability of the software package.
Experts, nonexperts, and small groups can participate
using software that they could not write on their own,
and new ideas can be tested with minimal overhead.
Since the current SN fitting models are based on externally 
performed training procedures that cannot be duplicated,
there is an urgent need to develop turn-key SN training programs. 
Scientists interested in studying SN light curves are invited to 
use and make contributions to {\tt SNANA}.

\clearpage
\bibliographystyle{apj}
\bibliography{snana}

  \end{document}